\documentclass[preprint]{aastex}

\newcommand{\myemail}{cwalsh13@qub.ac.uk}

\shorttitle{Molecular Anions in TMC-1}
\shortauthors{Walsh et al.}

\usepackage{wasysym}
\usepackage{subfigure}

\begin{document}

\title{The Effects of Molecular Anions on the Chemistry of Dark Clouds}

\author{Catherine Walsh\altaffilmark{1}, Nanase Harada\altaffilmark{2}, 
Eric Herbst\altaffilmark{3} and T. J. Millar\altaffilmark{1}}
\email{\myemail}

\altaffiltext{1}{Astrophysics Research Centre, School of Mathematics and Physics, 
Queen's University Belfast, University Road, Belfast, UK, BT7 1NN}
\altaffiltext{2}{Department of Physics, The Ohio State University, Columbus, 
OH 43210}
\altaffiltext{3}{Departments of Physics, Astronomy and Chemistry, 
The Ohio State University, Columbus, OH 43210}

\begin{abstract}
We have investigated the role of molecular anion chemistry 
in 
pseudo-time-dependent chemical models of dark clouds.   
With oxygen-rich elemental abundances, the addition of 
anions
results in a slight improvement in the overall 
agreement between model results and observations of 
molecular abundances in TMC-1 (CP). 
More importantly, with the inclusion of anions, 
we see an enhanced production efficiency of 
unsaturated carbon-chain neutral molecules, 
especially in the longer members of 
the families C$_{\mathrm{n}}$H, C$_{\mathrm{n}}$H$_2$, 
and HC$_{\mathrm{n}}$N.  
The use of carbon-rich elemental abundances in models 
of 
TMC-1 (CP) with anion chemistry worsens 
the agreement with observations obtained in the absence of anions.
\end{abstract}


\section{INTRODUCTION}\label{intro}

The possible existence of molecular anions (negatively charged ions) 
in the interstellar medium was discussed by \citet{dal73}, 
\citet{sar80}, and \citet{her81}.  
\citet{her81} calculated that the large electron affinity 
of carbon-chain molecules and hydrocarbon radicals would lead to 
efficient radiative electron attachment rate coefficients 
for species with more than 4-5 atoms
resulting in anion-to-neutral ratios on the order of a few percent.
Confirmation of this hypothesis was achieved following the measurement 
of the rotational spectrum of the molecular anion, C$_6$H$^-$, in the 
laboratory by \citet{mcc06}. 
This measurement allowed the verification, 
from existing astronomical spectra \citep{kaw95}, 
of the presence of C$_6$H$^-$ in the envelope 
of the carbon rich evolved star, IRC+10216, at 
an abundance 1\%-5\% that of the neutral molecule.
A successful astronomical search for C$_6$H$^-$ in 
TMC-1(CP) by \citet{mcc06} determined an anion-to-neutral 
ratio of $\sim$ 2.5\% in this source.

The rotational spectra of  C$_4$H$^-$, C$_8$H$^-$, 
and C$_3$N$^-$ have since been measured in the 
laboratory \citep{gup07,tha08} with subsequent 
detections of C$_4$H$^-$ \citep{cer07}, 
C$_8$H$^-$ \citep{rem07}, and C$_3$N$^-$ \citep{tha08} in 
the envelope of IRC+10216 and C$_8$H$^-$ in TMC-1 \citep{bru07}.  
Prompted by the observation of various carbon-chain molecules 
with large abundances in L1527, a class O/I protostar, by \citet{sak08}, 
C$_6$H$^-$ was searched for and confirmed to be present  
in this source \citep{sak07} followed closely by the detection of C$_4$H$^-$ 
\citep{agu08}.
A survey of galactic molecular sources by \citet{gup09} detected 
C$_6$H$^-$  in two futher sources, the pre-stellar cloud, L1544, 
and the protostellar object, L1521F, suggestive of the
ubiquitousness of molecular anions and C$_6$H$^-$ in particular.  
\citet{cer08} have attributed a series of rotational lines 
observed in the envelope of IRC+10216 to a new molecular anion, 
C$_5$N$^-$, although confirmation of this detection awaits successful 
measurement of the rotational spectrum of the molecule in the 
laboratory.

Following the identification of C$_6$H$^-$ in the envelope 
of IRC+10216 and the detection in TMC-1, \citet{mil07} 
expanded the current release of the UMIST Database for 
Astrochemistry \citep{woo07} and the existing chemical network 
for the envelope of IRC+10216 \citep{mil00} to include the molecular anions, 
C$_{\mathrm{n}}$$^{-}$ with n = 5 - 10 and C$_{\mathrm{n}}$H$^-$ with n = 4 - 10, 
using calculated rate coefficients for radiative electron attachment.  
 Their results were limited to calculated anionic 
and precursor neutral column densities.
For TMC-1 (CP), they found that the calculated `early time' 
C$_6$H$^-$ column density of $1.35\times10^{11}$ cm$^{-2}$
was very close to that measured by \citet{mcc06}, $1\times10^{11}$ cm$^{-2}$.
The calculated anion-to-neutral ratio of 5.2\% was also in good agreement 
with the observed value of 2.5\%.  More recently, \citet{har08} investigated
the chemistry of the protostellar object, L1527, concentrating on the formation and 
depletion of molecular anions.
\citet{wak08} considered the effects of negatively charged PAHs (polycyclic
aromatic hydrocarbons) on dense cloud chemistry and found that PAH$^-$ can replace
electrons as the dominant negative charge carrier, an effect which depends strongly
on the size of the PAHs considered.

The presence of molecular anions at significant abundances in the sources 
discussed above suggests that anions play a more significant role 
in the chemistry  of astrophysical environments than 
previously believed.   
In this paper, we investigate the effects that the inclusion of molecular 
anion 
chemistry has on the chemistry of dark clouds by
running simple pseudo-time-dependent models using 
the most recent release of the UMIST Database for Astrochemistry \citep{woo07}, 
henceforth referred to as `Rate06',
and the latest OSU (Ohio State University) network, osu.03.2008 
(see http://www.physics.ohio-state.edu/$\sim$eric/).  
 We compare our calculated fractional abundances for models both 
excluding and including anions with fractional abundances observed 
towards the cyanopolyyne (CP) peak of the
dark cloud core, TMC-1, to determine the role that 
these species play in the gas-phase chemistry of dark clouds. 
  
In Section \ref{chemmod}, we describe our anionic chemistry
and the initial chemical abundances and physical conditions adopted for our 
standard model.  
Section \ref{res} contains detailed results obtained with 
and without anions for both networks, discusses the effects of 
anions on the chemistry of selected classes of neutral species and
compares our results with observations towards the molecular-rich region, 
TMC-1 (CP). 
Section \ref{crich} shows how our results are affected by the use of 
carbon-rich elemental abundances.    
Finally, in Section \ref{disc} we summarize our main findings.

\section{CHEMICAL MODEL}\label{chemmod}

We have expanded both Rate06 and the OSU network to include the molecular anions 
C$_{\mathrm{n}}$$^-$ (n = 3 - 10), C$_{\mathrm{n}}$H$^-$ (n = 4 - 10), 
and C$_{\mathrm{n}}$N$^-$ (n = 3, 5), building upon the previous work by 
\citet{mil07} by increasing the number of molecular anions studied  
and expanding the reaction set for each anion to incorporate new 
reaction channels and rate coefficients from recently published experimental 
work \citep{eic07}.  
 We use the dipole-enhanced version of Rate06, which includes an enhancement of 
ion-neutral rate coefficients at low temperatures 
for reactions where the neutral has a large, permanent dipole 
moment ($>$ 0.9 Debye).  
The resulting rate coefficients have a T$^{-1/2}$ dependence at low 
temperatures \citep{her86,woo07}.  
A similar but not identical enhancement for the rate coefficients of 
ion-dipolar reactions is used in the OSU network \citep{her86}.  

The anions are formed primarily via radiative electron attachment,
\begin{eqnarray}
\mathrm{X} + e^- &\rightarrow& \mathrm{X}^- + h\nu,
\label{e-attach}
\end{eqnarray}
with updated attachment rate coefficients from recent theoretical calculations 
by \citet{her08}.
The hydrocarbon anions are destroyed through reactions with H, C, O, N, 
by photo-detachment, and by mutual neutralization with abundant cations 
such as C$^+$.
The destruction channels are listed in reactions (\ref{Hreact}) to (\ref{mutneut}) 
with m = 0 or 1:
\begin{eqnarray}
\mathrm{C_nH_m}^- + \mathrm{H} &\rightarrow& \mathrm{C_nH_{m+1}} + e^-,             \label{Hreact}\\
\mathrm{C_nH_m}^- + \mathrm{C} &\rightarrow& \mathrm{C_{n+1}H_{m}} + e^-,           \label{Creact}\\
\mathrm{C_nH_m}^- + \mathrm{O} &\rightarrow& \mathrm{CO} + \mathrm{C_{n-1}H_m}^-,   \label{Oreact}\\ 
\mathrm{C_nH_m}^- + \mathrm{N} &\rightarrow& \mathrm{CN} + \mathrm{C_{n-1}H_m}^-,   \label{Nreact}\\
                               &\rightarrow& \mathrm{CN}^- + \mathrm{C_{n-1}H_m},   \nonumber\\
                               &\rightarrow& \mathrm{C_3N}^- + \mathrm{C_{n-3}H_m}, \nonumber\\
                               &\rightarrow& \mathrm{C_5N}^- + \mathrm{C_{n-5}H_m}, \nonumber\\
                               &\rightarrow& \mathrm{H_mC_nN} + e^-,                \nonumber\\  
\mathrm{C_nH_m}^- + h\nu       &\rightarrow& \mathrm{C_nH_m} + e^-,                 \label{photreact}\\
\mathrm{C_nH_m}^- + \mathrm{X}^+ &\rightarrow& \mathrm{C_nH_m} + \mathrm{X}.        \label{mutneut}
\end{eqnarray}
We have taken the rate coefficients for reaction (\ref{Hreact}) from \citet{bar01} 
and for reactions (\ref{Creact}), (\ref{Oreact}), and (\ref{Nreact}) 
from \citet{eic07} for the species for which they were measured.
In the absence of laboratory data, we have used estimated rate coefficients 
based on the experimental measurements.
The rate coefficients and reaction channels for reaction (\ref{Nreact}) 
are particularly uncertain because \citet{eic07} were unable to detect 
associative detachment as a reaction channel, 
a potentially important synthetic route to the HC$_{\mathrm{n}}$N molecules.  
We discuss the implications of these uncertainties in \S\ref{hcnn}.   
The photo-detachment rate coefficients (reaction [\ref{photreact}]) adopted are those listed in 
Table 1 of 
\citet{mil07}, although in our model, this is an unimportant destruction 
mechanism as TMC-1 is well-shielded from external sources of UV radiation.
The mutual neutralization reactions (reaction [\ref{mutneut}]) each have a rate 
coefficient $k = 7.5 \times 10^{-8}(T/300)^{-0.5}$ cm$^3$ s$^{-1}$ as 
included by \citet{har08} in their chemical model of L1527. 
We have assumed in all instances where appropriate that the products are 
obtained by a simple electron transfer.
The C$_3$N$^-$ chemistry we have added includes that listed in Table 1 
of \citet{har08} plus formation routes via the molecular anions, 
C$_{\mathrm{n}}$$^-$ and C$_{\mathrm{n}}$H$^-$ 
(eq. [\ref{Nreact}]), as observed in the experimental results of \citet{eic07}.
We have included C$_5$N$^-$ using the same destruction reactions and rate 
coefficients as for C$_3$N$^-$ with a formation reaction rate coefficient 
(reaction [\ref{e-attach}]) calculated by E. Herbst (private communication) 
using the method of \citet{her08}, to be $1.25\times10^{-7}(T/300)^{-0.5}$ 
cm$^3$ s$^{-1}$.  
Due to the inclusion of C$_{10}$$^-$ and C$_{10}$H$^-$, we extrapolated both 
networks to include further hydrocarbon species containing 10 carbon atoms hence 
restricting the hydrocarbon chemistry to molecules with 11 carbon atoms or less. 

We model the gas-phase chemistry of a dark cloud core 
by treating the source 
as an homogeneous, isotropic cloud with the constant physical parameters, 
n(H$_2$) = 10$^4$ cm$^{-3}$, T = 10 K, and A$_{\mathrm{v}}$ = 10 mag, 
using a cosmic-ray ionisation rate of $1.3 \times 10^{-17}$ s$^{-1}$, 
and performing pseudo-time-dependent calculations, in which the 
chemistry 
evolves from initial abundances that are atomic except for molecular hydrogen, 
which is produced efficiently at an earlier stage. 
We employ the commonly used oxygen-rich low-metal elemental abundances, 
as listed in Table \ref{table1} and suggested originally by \citet{gra82}. 
Because pseudo-time-dependent gas-phase models of dark clouds are 
relatively predictive compared with more complex physical models,  
using such a simplified model helps us to accurately identify particular 
reactions or, more commonly, systems of reactions that are important and 
require further study.    
 
Any comparison between our modelled results and  observations in TMC-1 (CP) 
could, of course, be compromised by our use of a simplified model for a dark cloud.  
We would expect that employing a more detailed physical model could 
lead to different results from those presented here;
however, such models are certainly not unique and therefore bring their own 
limitations.  
For example, recent work by \citet{has09}, in which a chemical dynamical-model
following both the gas-phase and surface chemistry as a dense cloud forms via a 
shock, finds little difference in the gas-phase abundances of large molecules 
although the time-scales for formation are certainly different.
Chemical factors, such as our neglect of accretion of gas-phase species 
onto dust grains, will also affect gas-phase abundances,  
although the timescales we are discussing here, 
$\sim 10^5$ yr, are only just at the stage where accretion
becomes important for the conditions we have adopted. 

\section{RESULTS}\label{res}

Table~\ref{table2} shows our calculated  abundances using both networks for an 
assortment of molecules at the times of best agreement 
(typically $1-2 \times 10^{5}$ yr) for TMC-1 (CP)
along with the observed values or upper limits.  
Model results are shown for calculations with and without anions.
As can be seen, there is little difference between the results 
obtained with Rate06 and with the OSU network.  
We first discuss the anions and then consider families of 
neutral species strongly affected by the inclusion of anions.  
Throughout this section, we display the results from the models using
Rate06 only because the results from the models using the OSU network are
similar.  

\subsection{Anions} \label{anions}

\begin{figure}
\figurenum{1}
\centering
\subfigure
	{\includegraphics[angle=270,scale=0.4]{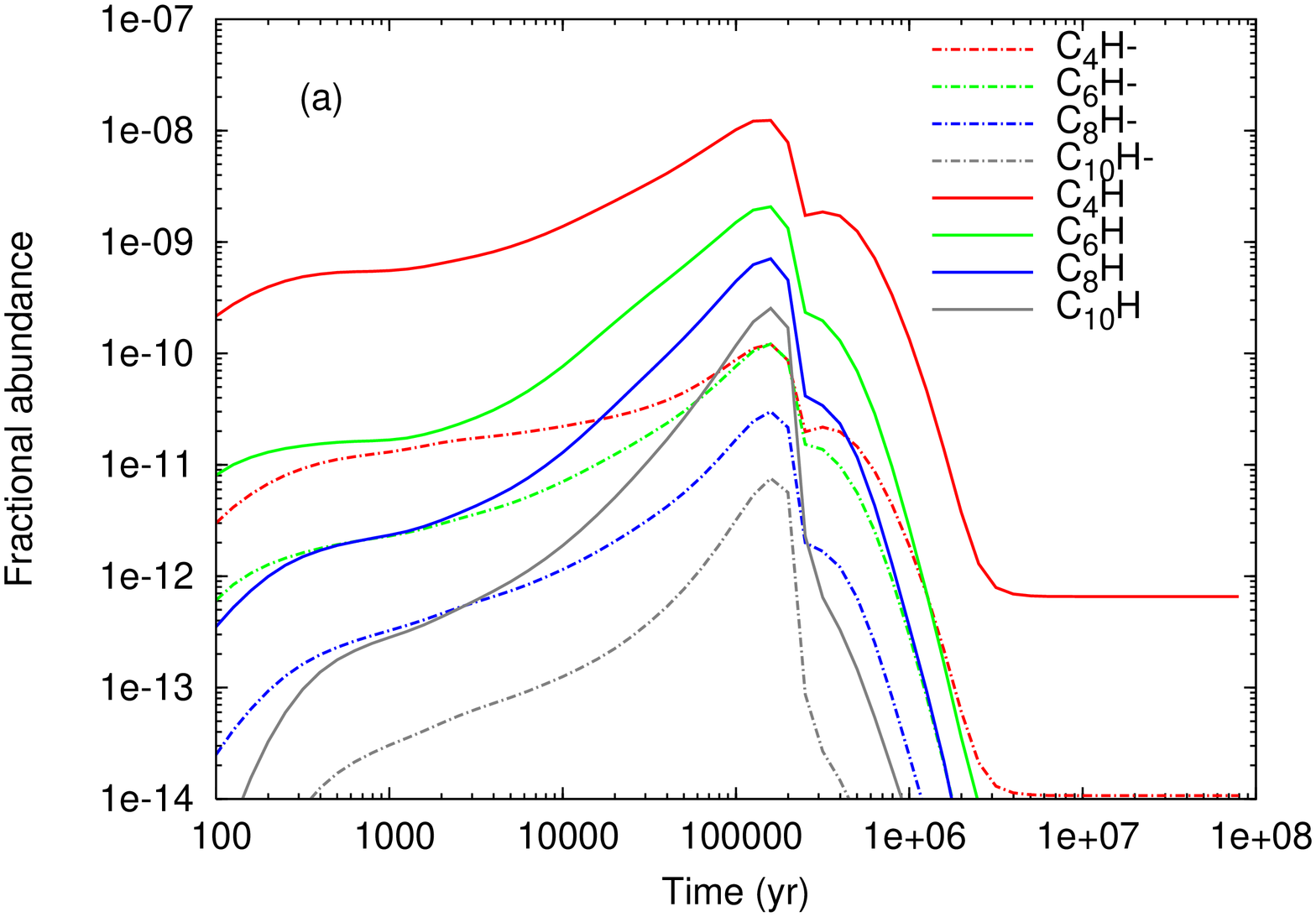}}	
        {\includegraphics[angle=270,scale=0.4]{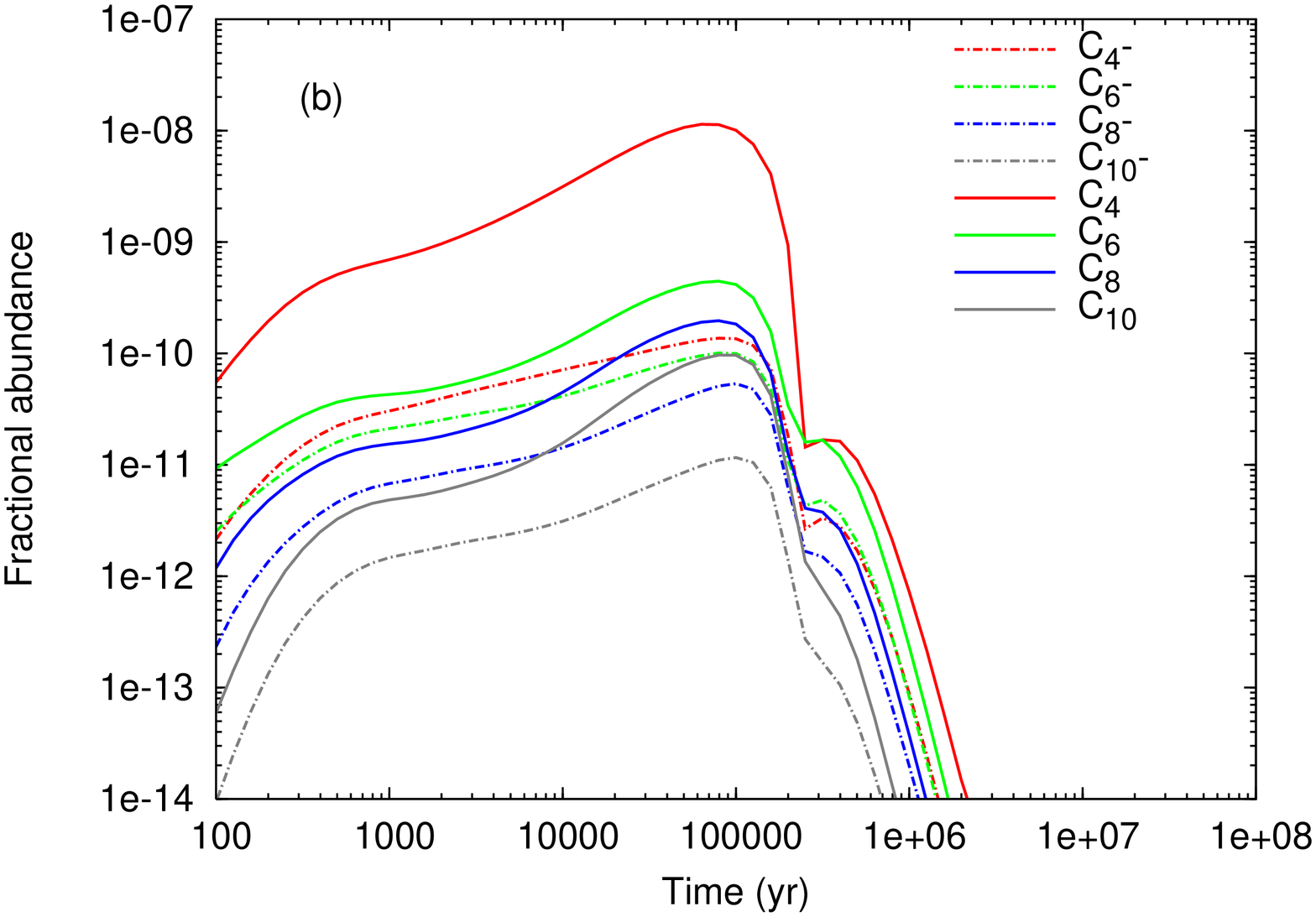}}        
        {\includegraphics[angle=270,scale=0.4]{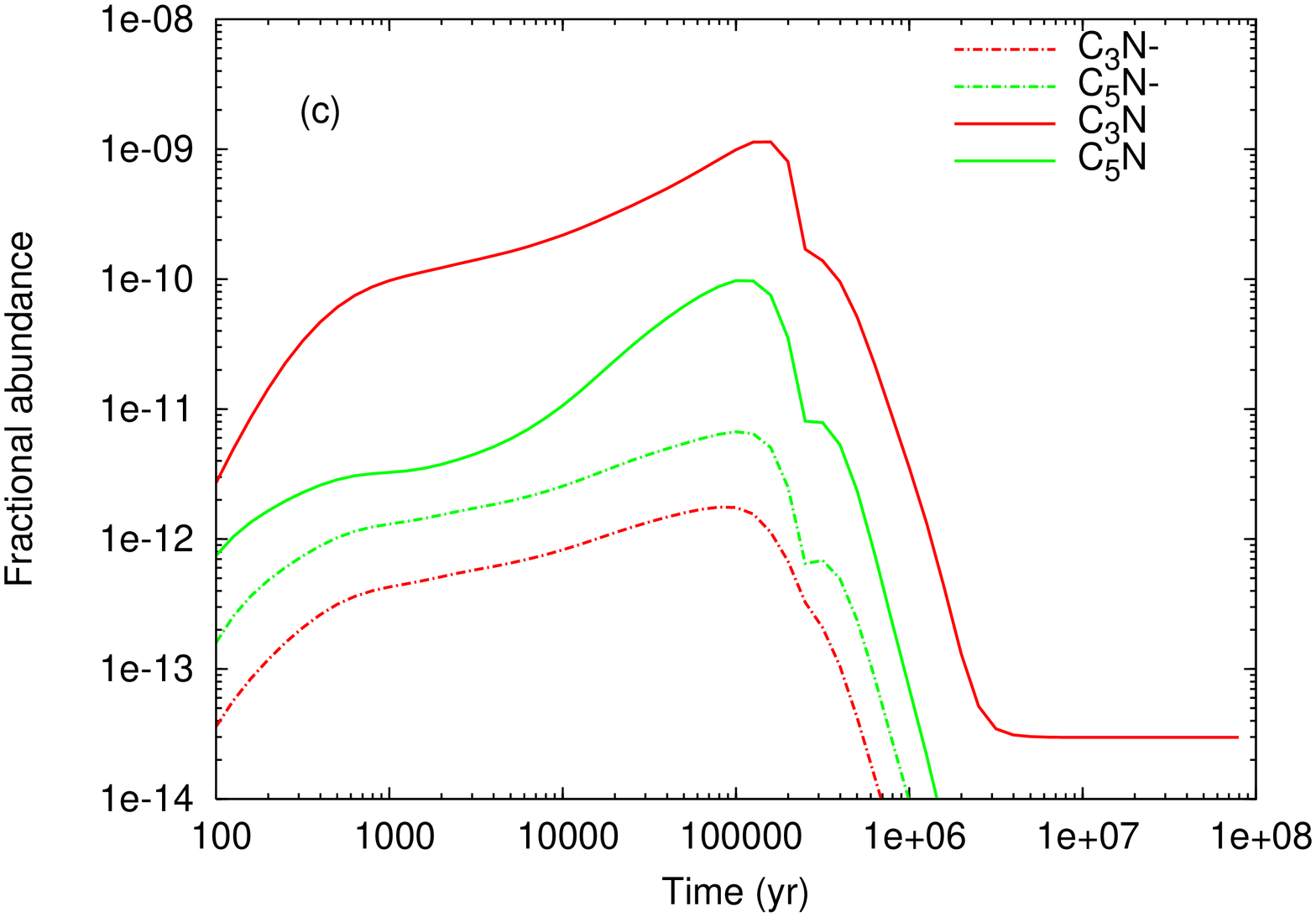}}        
\caption{The abundances with respect to H$_{2}$ density of (a) C$_{\mathrm{n}}$H$^-$ 
for n = 4, 6, 8, and 10, (b) C$_{\mathrm{n}}^-$ for n = 4, 6, 8, 10, 
and (c) C$_{\mathrm{n}}$N$^-$ for n = 3, 5 and their corresponding neutral
analogs, 
as a function of time.}
\label{figure1}
\end{figure}

The calculated fractional abundances of the molecular anions, 
C$_\mathrm{n}$H$^-$ for n = 4, 6, 8, and 10, 
C$_\mathrm{n}$$^-$ for n = 4, 6, 8, and 10, and C$_\mathrm{n}$N$^-$ 
for n = 3 and 5, along with the neutral precursor or each anion, 
are plotted as a function of time in Figures~\ref{figure1}~(a), 
\ref{figure1}~(b) and \ref{figure1}~(c), 
respectively.  
Because the main route of formation of the anions at all times is radiative 
electron attachment, the abundances of the anions follow that of their 
neutral precursors with both generally inversely proportional to 
molecular size.
An exception is the abundance of C$_4$H$^-$, which at $\approx 10^5$ yr 
is comparable to that of C$_6$H$^-$ because of 
the slow radiative electron attachment rate coefficient 
calculated for C$_4$H \citep{her08}.
The larger abundance of C$_5$N$^-$ compared with C$_3$N$^-$ 
at all times, despite the smaller abundance of C$_5$N compared with C$_3$N,
is due to the much larger electron attachment rate coefficient calculated 
for C$_5$N ($\sim 10^{-7}$ cm$^{3}$ s$^{-1}$ compared
with $\sim 10^{-10}$ cm$^{3}$ s$^{-1}$). 
Although the reaction rate coefficient 
for the formation of C$_3$N$^-$ via dissociative attachment of HNC$_3$ is
two orders of magnitude larger than that of the radiative electron attachment 
(see Table 1 of \citet{har08}) the abundance of HNC$_3$ only reaches 
$\sim$ 0.01 that of C$_3$N after $\sim 10^5$ yr when dissociative attachment 
becomes comparable with radiative attachment.   
The abundances of both the anions and neutrals peak around $10^5$ yr and  
the destruction of the anions is dominated by reactions with H, C, O
and C$^+$ (reactions [\ref{Hreact}], [\ref{Creact}], 
[\ref{Oreact}], and [\ref{mutneut}], respectively).

In the absence of anions, 
the main destruction channels for C$_{\mathrm{n}}$H are via reactions with 
C, O, and N:
\begin{eqnarray}
\mathrm{C} + \mathrm{C_nH} & \rightarrow & \mathrm{C_{n+1}} + \mathrm{H}, \\
\mathrm{O} + \mathrm{C_nH} & \rightarrow & \mathrm{C_{n-1}H} + \mathrm{CO}, \\
\mathrm{N} + \mathrm{C_nH} & \rightarrow & \mathrm{C_{n-1}H} + \mathrm{CN}.
\end{eqnarray}
The inclusion of anions introduces radiative electron attachment 
(reaction [\ref{e-attach}]) as a destruction mechanism and it is this channel 
that dominates the destruction of the C$_{\mathrm{n}}$H molecules at all times, 
excepting C$_4$H for which the destruction via the reaction with O is comparable.

Anionic chemistry also introduces new formation routes for the 
C$_{\mathrm{n}}$H radicals.  
Ordinarily, these radicals are formed primarily via dissociative 
electron attachment to
hydrocarbon cations as well as via carbon insertion reactions  
(in reaction [\ref{elecattach}] m = 2 or 3):
\begin{eqnarray}
\mathrm{C_nH_m}^+ + e^-            & \rightarrow & \mathrm{C_nH} + \mathrm{H_{m-1}}\label{elecattach}, \\
\mathrm{C_{n+1}H}^+ + e^-          & \rightarrow & \mathrm{C_nH} + \mathrm{C}, \\
\mathrm{C} + \mathrm{C_{n-1}H_{2}} & \rightarrow & \mathrm{C_nH} + \mathrm{H}.
\end{eqnarray}
New formation routes become important past $\sim 10^3$ yr when appreciable
abundances of molecular anions have built up and C$_{\mathrm{n}}$H is formed 
via associative
electron detachment from carbon-chain anion and hydrocarbon anion precursors 
(see reactions [\ref{Hreact}] and [\ref{Creact}]):
\begin{eqnarray}
\mathrm{H} + \mathrm{C_n}^-      & \rightarrow & \mathrm{C_nH} + e^-\label{assdetach1},\\
\mathrm{C} + \mathrm{C_{n-1}H}^- & \rightarrow & \mathrm{C_nH} + e^-\label{assdetach2}.
\end{eqnarray}

Radiative electron attachment of the bare carbon-chain neutrals, C$_{\mathrm{n}}$, as a
destruction mechanism, only dominates prior to $10^4$ yr, at which
point reactions with O become comparable to this channel.
As with the hydrocarbon radicals, the inclusion of bare carbon-chain anions opens up an
additional formation route to the bare carbon-chain neutrals via associative electron
detachment (reaction [\ref{Creact}]):
\begin{eqnarray}
\mathrm{C_{n-1}}^{-} + \mathrm{C} &\rightarrow& \mathrm{C_{n}} + e^{-}.
\label{assdetach3}
\end{eqnarray}

As can be seen in Table~\ref{table2}, 
the abundances for C$_6$H$^-$ and C$_8$H$^-$ are over-predicted 
at the time of best overall agreement by around 
an order of magnitude, with the abundance for C$_4$H$^-$ around 
50 times greater than the estimated upper limit by \citet{tha08}.  
There is a sharp decline in the predicted C$_{\mathrm n}$H$^{-}$ abundances 
at somewhat later times, however, as can be seen in Figure~\ref{figure1}~(a).  
The abundance of C$_4$H$^-$ only falls below the upper limit estimated by 
\citet{tha08} after $\sim$ 10$^6$ yr indicating that our electron attachment 
rate for C$_4$H is too large or that we are neglecting further destruction 
mechanisms for this species.
In contrast to the anions already discussed, 
the predicted abundance of C$_{3}$N$^-$ at the time of 
best overall agreement is in accordance with the upper limit determined by 
\citet{tha08}.  
  
\begin{figure}
\figurenum{2}
\includegraphics[angle=270,scale=0.6]{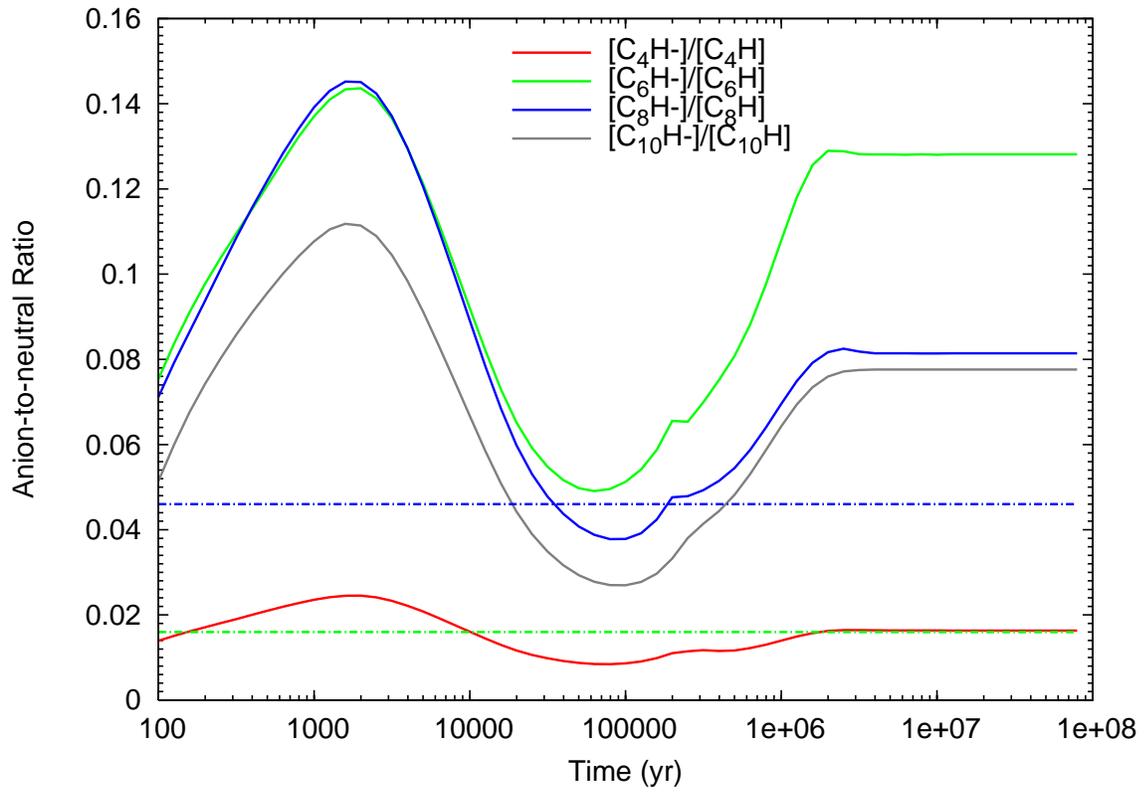}
\caption{The anion-to-neutral ratio C$_{\mathrm{n}}$H$^-$/C$_{\mathrm{n}}$H 
for n = 4, 6, 8, and 10, as a 
function of time.  The dotted-dashed horizontal lines are the observed ratios 
for C$_{6}$H$^{-}$ and C$_{8}$H$^{-}$.}
\label{figure2}
\end{figure}

The calculated anion-to-neutral ratio, C$_{\mathrm{n}}$H$^{-}$/C$_{\mathrm{n}}$H 
for n = 4, 6, 8, and 10  is plotted as a function of time in Figure~\ref{figure2}, 
along with the observed ratios for C$_6$H and C$_8$H.
The calculated ratio for all species reaches a minimum around 10$^5$ yr, the same
time that the peak in individual fractional abundances is reached.  
The model results for Rate06 yield 5.2\% for the C$_6$H ratio and 3.9\% 
for the C$_8$H ratio.  
These compare with the observed values for C$_6$H and C$_8$H of 1.6\% and 
4.6\%, respectively. 

\subsection{Effects on Other Species} \label{otherspecs}

The main aim of this paper is to investigate the effects that the
inclusion of anions has on the abundances and evolutionary behavior of other 
species in established chemical models.  
We have identified several families of species that are affected; these
include the neutral precursors of the anions, C$_{\mathrm{n}}$, C$_{\mathrm{n}}$H, 
and C$_{\mathrm{n}}$N as well as the families of molecules 
C$_{\mathrm{n}}$H$_2$ and HC$_{\mathrm{n}}$N.  
 The sensitivity of species in these families to the inclusion of anions can be 
seen in Table \ref{table2} for both networks.

\subsubsection{C$_n$} \label{cn}

Figure~\ref{figure3}~(a) displays the fractional abundances of the neutral 
C$_{\mathrm{n}}$ molecules for n = 4, 6, 8, and 10 as a function of
time for the models with and without anions.  
It is immediately apparent that the abundances of C$_6$, C$_8$, and C$_{10}$ 
are enhanced at all times.
The relative enhancement is more significant with increasing molecular size 
with little change in the abundance of C$_4$, in contrast to that of
C$_{10}$, which increases by around 2 orders of magnitude 
in the presence of anions (see Table \ref{table2}).  
As discussed previously in \S\ref{anions}, 
introducing anions to the chemistry includes a new formation route via 
associative electron detachment (reaction [\ref{assdetach3}]) 
whereby larger neutral carbon-chains can build up from smaller anionic precursors 
through repeated insertion of a carbon atom.  

\subsubsection{C$_n$H} \label{cnh}

\begin{figure}
\figurenum{3}
\centering
\subfigure
	{\includegraphics[angle=270,scale=0.4]{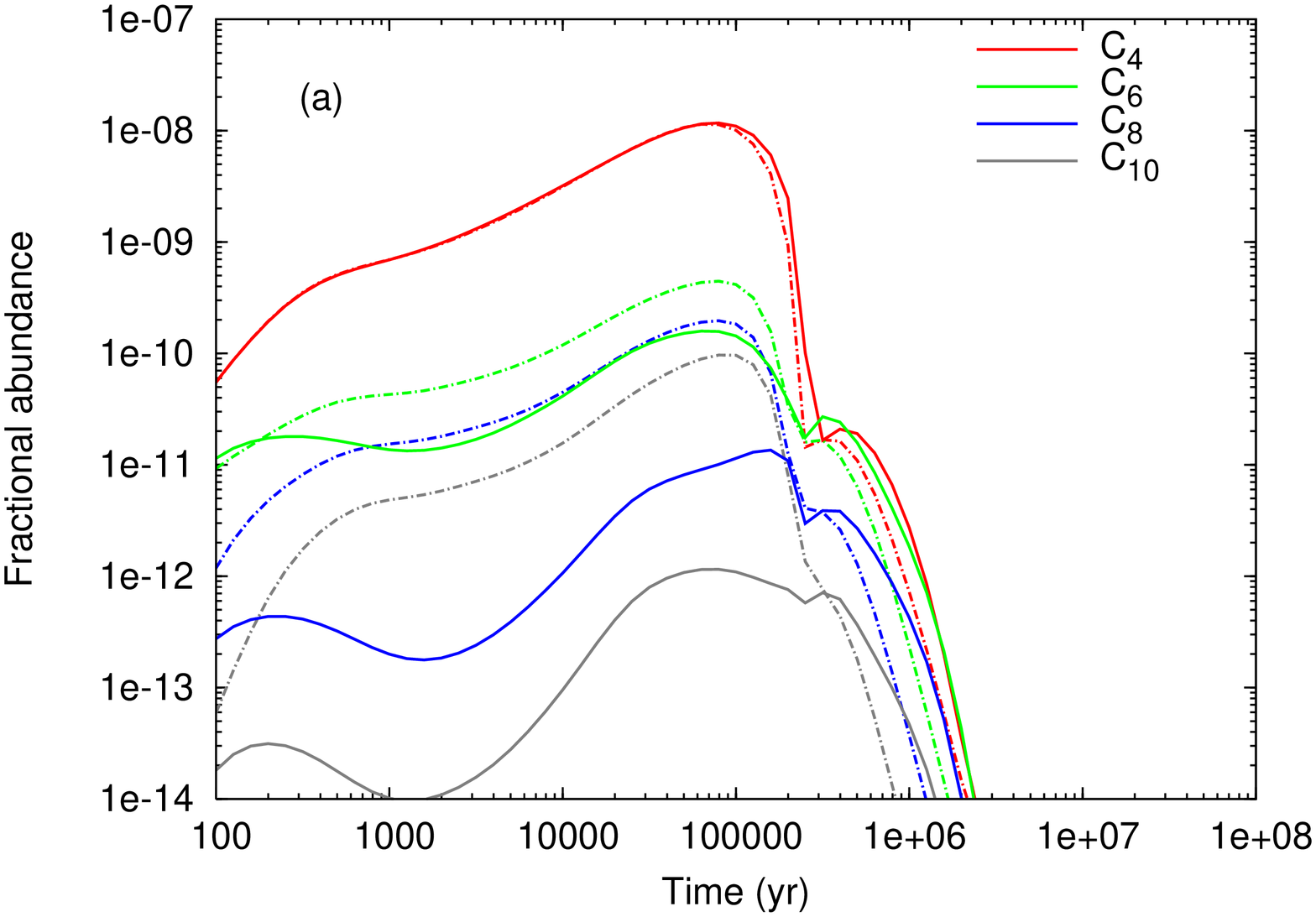}}	
        {\includegraphics[angle=270,scale=0.4]{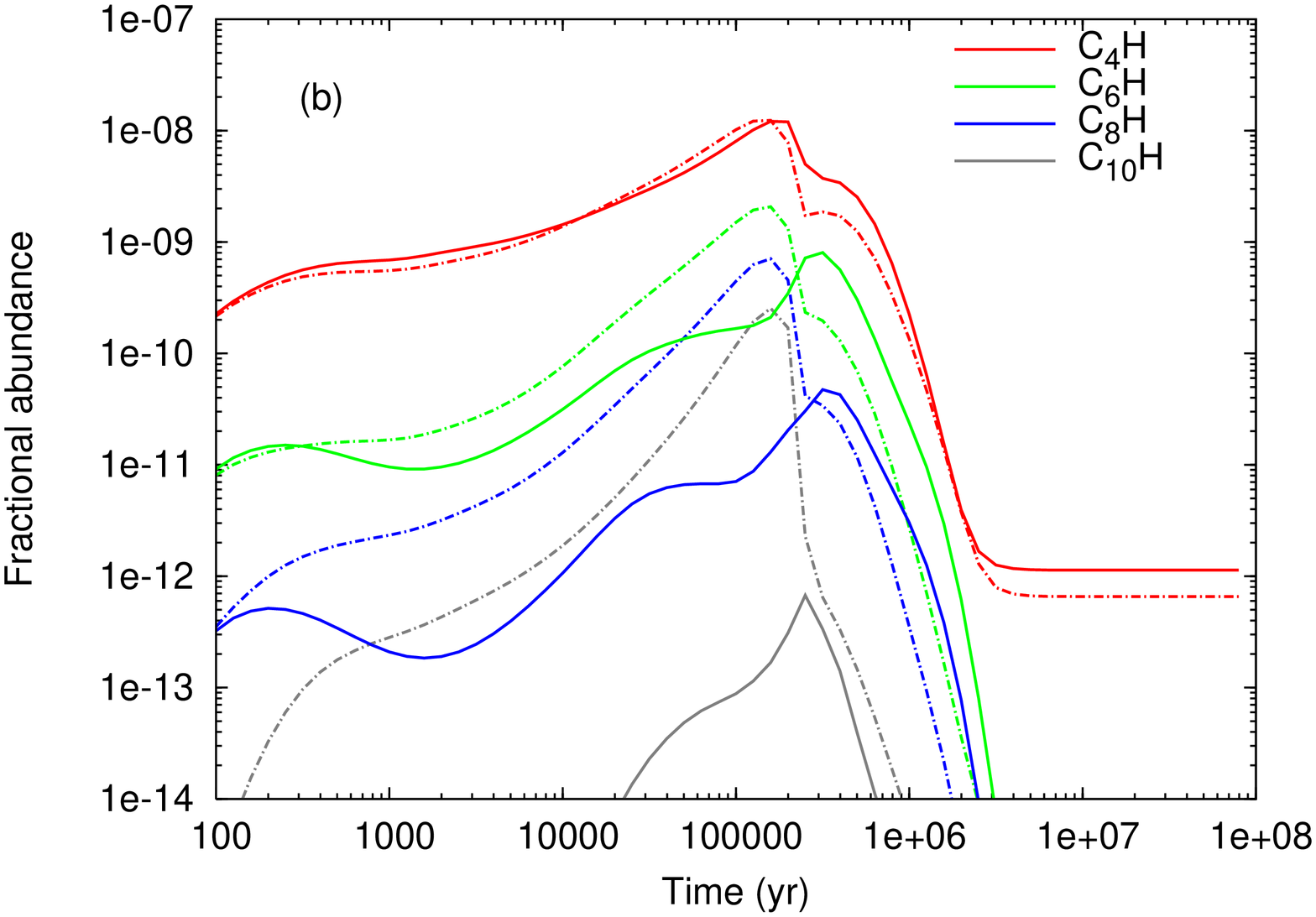}}        
        {\includegraphics[angle=270,scale=0.4]{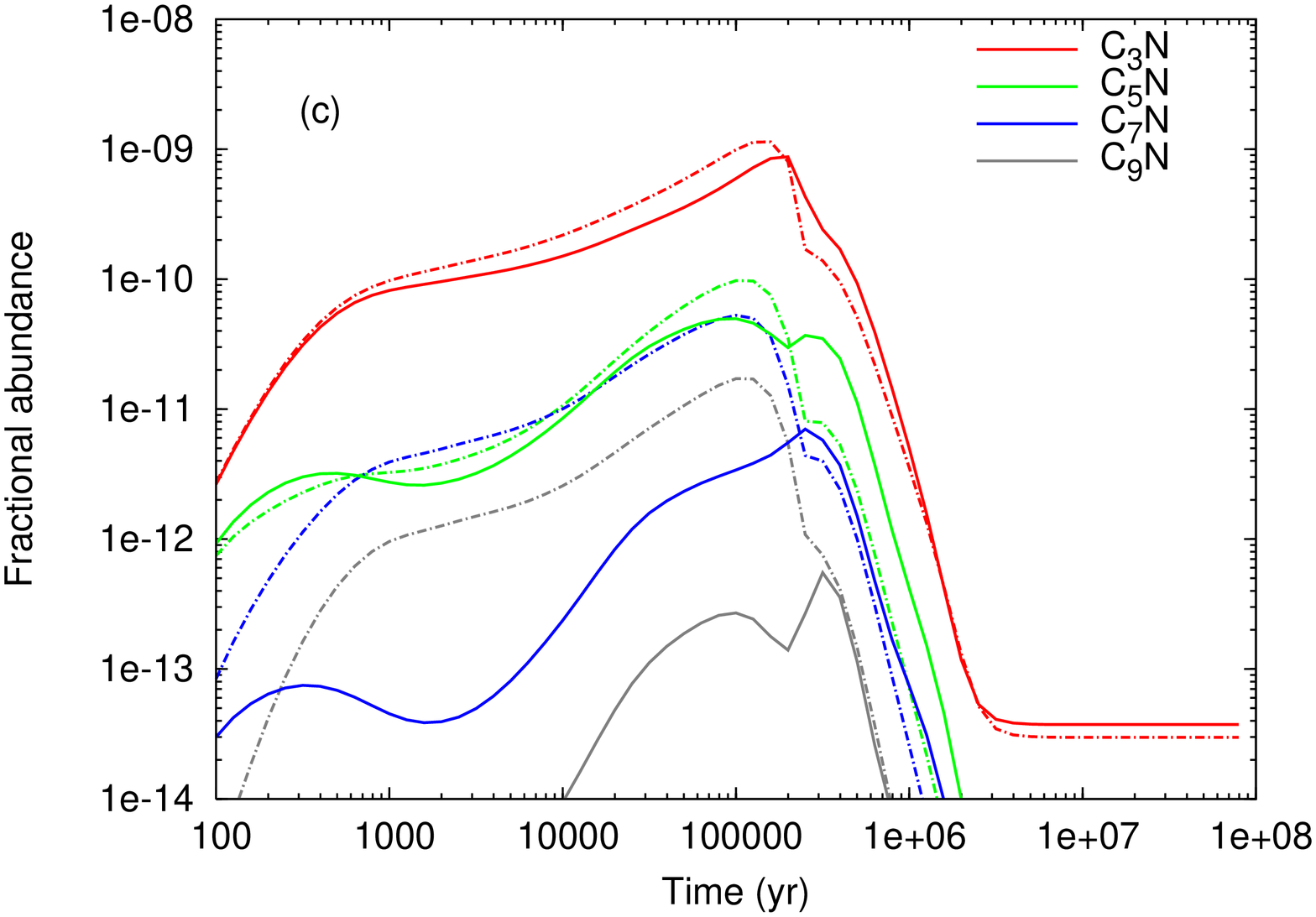}}        
\caption{The abundances with respect to H$_{2}$ density of (a) C$_{\mathrm{n}}$ 
for n = 4, 6, 8, and 10, (b) C$_{\mathrm{n}}$H for n = 4, 6, 8, and 10, 
and (c) C$_{\mathrm{n}}$N for n = 3, 5, 7, and 9, as a function 
of time, with (dotted-dashed lines) and without (solid lines) anions.}
\label{figure3}
\end{figure}

Figure~\ref{figure3}~(b) shows the fractional abundances of the C$_{\mathrm{n}}$H 
molecules for the models with and without anions.  
As in Figure~\ref{figure3}~(a) for C$_{\mathrm{n}}$, there is an obvious 
enhancement of the larger hydrocarbon radicals,
C$_6$H, C$_8$H, and C$_{10}$H, at all times prior to $10^6$ yr  
with the enhancement increasing with molecular size.  
The addition of anions introduces two new formation mechanisms via 
associative detachment from 
bare carbon-chain anions (reaction [\ref{assdetach1}]) and hydrocarbon anions 
(reaction [\ref{assdetach2}]), again
introducing a new route to the 
formation of larger hydrocarbon radicals from smaller ones.  
At the time of overall best agreement, as seen in Table \ref{table2}, 
 the inclusion of anions enhances the abundance of C$_6$H from a
value which is $\approx$ 3 times smaller than observed to one $\approx$ 
3 times larger than observed, while the abundance of 
C$_8$H is enhanced from a value $\approx$ 4 times smaller to one 
$\approx$ 14 times larger than observed.  
There is no enhancement of the C$_4$H  abundance, 
which remains a factor of $\approx$ 5 below the observed value.  
As in the case for the hydrocarbon anions, the 
abundances of both C$_6$H and C$_8$H
reach their best agreement with observation
at slightly later times, while that of C$_4$H diminishes further.

\subsubsection{C$_n$N} \label{cnn}

The fractional abundances of the C$_{\mathrm{n}}$N
radicals, with n = 3, 5, 7, and 9, are plotted in Figure~\ref{figure3}~(c), 
as a function of time.  
The abundances of all of these species are enhanced in the model with anions, 
compared with the results for the model without anions, prior to 
$\sim 3 \times 10^{5}$ yr.
As with the carbon chains and hydrocarbons, the biggest relative enhancement 
is seen in the larger species, C$_7$N and C$_9$N, with the peak fractional abundance 
of C$_9$N increasing by around 2 orders of magnitude from $\sim 10^{-13}$ 
to $\sim 10^{-11}$.  
When anions are not included in the model, the main routes to the formation of 
C$_{\mathrm{n}}$N for n = 5, 7, and 9 are  
\begin{eqnarray}
\mathrm{H_2C_nN}^+ + e^-   & \rightarrow & \mathrm{C_nN} + \mathrm{H_2}, \label{CnNform1}\\
\mathrm{H_3C_nN}^+ + e^-   & \rightarrow & \mathrm{C_nN} + \mathrm{H_2} + \mathrm{H}, \label{CnNform2}\\
\mathrm{N} + \mathrm{C_nH} & \rightarrow & \mathrm{C_nN} + \mathrm{H}, \label{CnNform3}
\end{eqnarray}
with destruction via reactions with O and N atoms.
Reaction (\ref{CnNform1}) dominates the formation up to $\sim 10^5$ yr after 
which reaction (\ref{CnNform3}) is comparable to or dominates over reaction 
(\ref{CnNform1}).
For C$_3$N, reaction (\ref{CnNform3}) is the main route of formation at all times.  

The inclusion of anions introduces a new formation route via the bare carbon chain 
anions, C$_{\mathrm{n}}$$^-$:
\begin{eqnarray}
\mathrm{N} + \mathrm{C_n}^- & \rightarrow & \mathrm{C_nN} + e^-,\label{CnNform4}
\end{eqnarray}
and it is this route that dominates the formation for n = 5, 7, and 9 up to 
$\sim 10^5$ yr after which reactions (\ref{CnNform1}) and (\ref{CnNform3}) 
take over.  
As in the case without anions, the main formation route for C$_3$N is through 
reaction (\ref{CnNform3}).  
As the abundances of the C$_{\mathrm{n}}$H radicals are enhanced with the inclusion 
of anions this also increases the influence of reaction (\ref{CnNform3}) over that of  
reaction (\ref{CnNform1}) in contrast to the case without anions.

\subsubsection{C$_n$H$_2$} \label{cnh2}

Whilst the effect on the abundances of C$_\mathrm{n}$ and C$_\mathrm{n}$H due 
to the inclusion of their respective anions may come as little surprise, 
other families of species are also affected.
Figure~\ref{figure4}~(a) shows the time-dependent fractional abundance of 
C$_\mathrm{n}$H$_2$ for n = 4, 6, 8, and 10.  
The abundances of all species are enhanced at all times in the model with 
anions included, with a more marked effect for C$_8$H$_2$ and C$_{10}$H$_2$.
In the absence of anions, the C$_\mathrm{n}$H$_2$ species form via 
dissociative recombination, e.g. reactions (\ref{dissattach1}) and 
(\ref{dissattach2}), or via neutral-neutral reactions,
\begin{eqnarray}
\mathrm{C_nH_3}^{+} + e^-           & \rightarrow & \mathrm{C_nH_2} + \mathrm{H}, \label{dissattach1}\\
\mathrm{C_nH_4}^{+} + e^-           & \rightarrow & \mathrm{C_nH_2} + \mathrm{H_2}, \label{dissattach2}\\
\mathrm{C_{n-2}H_2} + \mathrm{C_2H} & \rightarrow & \mathrm{C_nH_2} + \mathrm{H}. \label{neutneut}
\end{eqnarray}
Adding anions to the chemistry opens a new synthetic channel via associative 
detachment of the hydrocarbon anions, C$_{\mathrm{n}}$H$^-$, with H atoms 
(reaction [\ref{Hreact}]) and it is this channel which dominates up to $\sim
10^6$ yr.
The C$_{\mathrm{n}}$H$_2$ species in the dominant polyyne isomer, 
HC$_{\mathrm{n}}$H, are not observable via rotational spectroscopy because this 
isomer has zero dipole moment.  
The cumulene carbene forms H$_2$C$_{\mathrm n}$ are detectable but in general 
cannot be formed by associative detachment because the reactions are endothermic 
\citep{her08}.  

\subsubsection{HC$_n$N} \label{hcnn}

\begin{figure}
\figurenum{4}
\centering
\subfigure
	{\includegraphics[angle=270,scale=0.5]{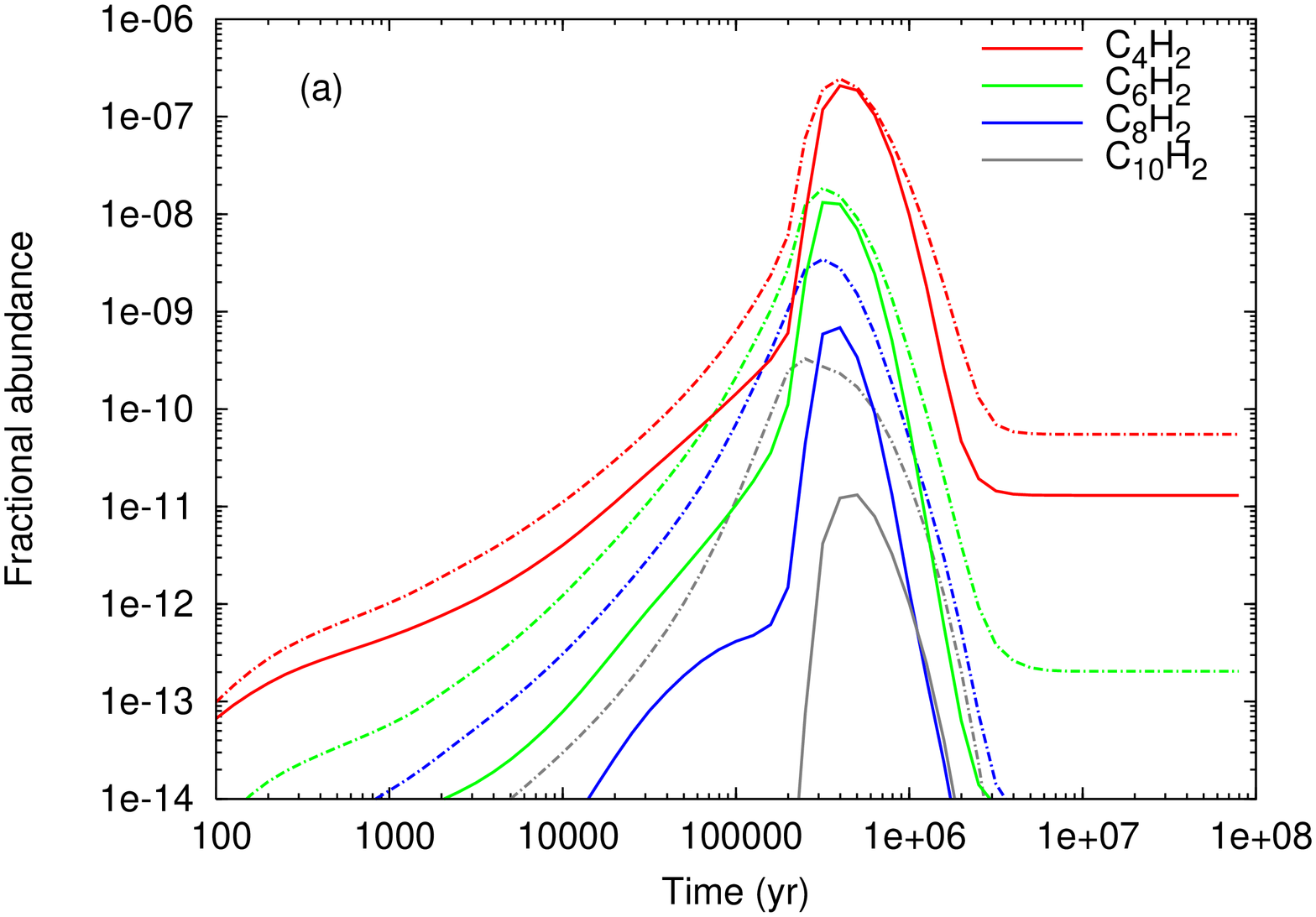}}	
        {\includegraphics[angle=270,scale=0.5]{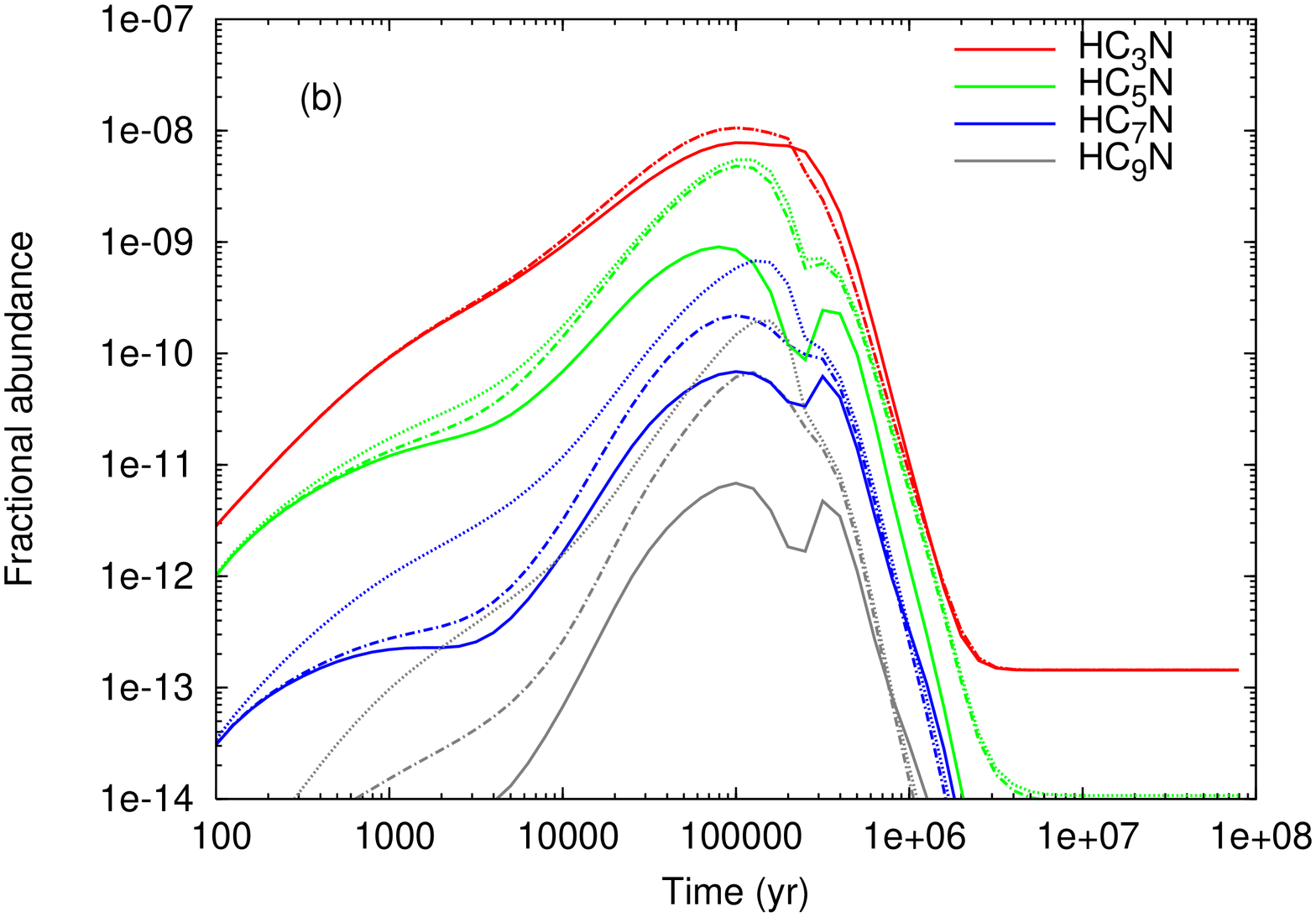}}        
\caption{The abundances with respect to H$_{2}$ density of (a)
C$_{\mathrm{n}}$H$_2$ 
for n = 4, 6, 8, and 10, and (b) HC$_{\mathrm{n}}$N for n = 3, 5, 7, and 9, 
as a function of time, with (dotted-dashed lines) and without (solid lines) anions.  
The dotted lines in the HC$_{\mathrm{n}}$N plot are results from the model using the 
larger rate coefficients for $\mathrm{C_nH^- + N} \rightarrow \mathrm{HC_nN + e^-}$.}
\label{figure4}
\end{figure}

Figure~\ref{figure4}~(b) shows the abundances
of the HC$_{\mathrm{n}}$N molecules for n = 3, 5, 7, and 9 and again we see a 
marked enhancement 
in the abundances of all
species when anions are included in the networks.  
The synthesis of the larger cyanopolyyne molecules proceeds (as for many
neutral molecules) via the dissociative recombination of a 
cationic molecular precursor:
\begin{eqnarray}
\mathrm{H_2C_nN}^+ + e^- &\rightarrow& \mathrm{HC_nN} + \mathrm{H}, \\
\mathrm{H_3C_nN}^+ + e^- &\rightarrow& \mathrm{HC_nN} + \mathrm{2H},
\end{eqnarray}
with a secondary route through the reaction
\begin{eqnarray}
\mathrm{C_{n-1}H_2} + \mathrm{CN} &\rightarrow& \mathrm{HC_nN} + \mathrm{H}.
\label{HCnNform3}
\end{eqnarray}
The addition of anions to the chemistry enhances the abundances of the
C$_\mathrm{n}$H$_2$ species which increases the effect of reaction
(\ref{HCnNform3}).  
There is also a more subtle influence on the abundances of
H$_2$C$_\mathrm{n}$N$^+$ and H$_3$C$_\mathrm{n}$N$^+$ which are produced via a
series of different reaction pathways originating from hydrocarbon cations,
C$_\mathrm{n}$H$_\mathrm{m}$$^+$, the abundances of which are enhanced in the
presence of molecular anions.
Examples of this type of reaction are
\begin{eqnarray}
\mathrm{C_4H_2}^+ + \mathrm{HC_{n-4}N} &\rightarrow& \mathrm{H_3C_nN^+} + h\nu, \\
\mathrm{C_2H_2}^+ + \mathrm{HC_{n-2}N} &\rightarrow& \mathrm{H_3C_nN^+} + h\nu.
\end{eqnarray}
In addition, the inclusion of C$_3$N$^-$ and C$_5$N$^-$ introduces new formation 
mechanisms for HC$_3$N and HC$_5$N via associative electron detachment of each 
anion with H atoms.  
This reaction has a fast rate coefficient, on the order of 
10$^{-9}$ cm$^3$ s$^{-1}$, which has a large effect on the abundance of HC$_5$N, 
in particular.
 
The destruction of the hydrocarbon anions C$_\mathrm{n}$H$^{-}$ via reactions 
with atomic N (reaction [\ref{Nreact}]) has an associative detachment channel, 
which leads to the cyanopolyyne molecules for n = 3, 5, 7, and 9.
The rate coefficients for these reactions have been derived from experimental data 
by \citet{eic07} who found that the reactions of anions with an even number of 
carbon atoms proceed slowly, with rate coefficients on the order
of 10$^{-11}$ cm$^3$ s$^{-1}$.  
Branching ratios for the products were not determined, 
although contact with the authors led to their agreement 
that the associative detachment channel to form the 
cyanopolyynes could have a branching fraction of $\approx$ 0.5.  
For reactions with odd numbers of carbons in the hydrocarbon anions, 
i.e. n = 3, 5, 7, and 9, 
we have assumed that the reactions proceed with a rate coefficient the same 
order of magnitude, with an associative detachment channel that has a 
branching fraction 50\% of the total rate.
Inclusion of this reaction channel has little effect on the HC$_{\mathrm{n}}$N 
abundances due to the small rate coefficient used; however, 
we have also run a model with this reaction channel proceeding using a larger
rate coefficient of $10^{-10}$ cm$^3$ s$^{-1}$, on the basis that the 
reactions between C$_{\mathrm{n}}$H$^-$ species with odd values of n and 
atomic nitrogen are spin allowed.  
The results from this ``third'' model are also shown in Figure~\ref{figure4}~(b) 
(dotted lines) in addition to the results from the model without anions 
(solid lines) and our standard model with anions (dotted-dashed lines).
The associative detachment  channel futher enhances the abundances of 
HC$_7$N and HC$_9$N in particular, two molecules for which existing 
chemical networks have difficulty reproducing the observed abundances.
At the time of best agreement in the third  model ($1.6 \times 10^5$ yr), 
the abundances are in vastly better agreement with the observed abundances 
listed in Table \ref{table2} although they still lie somewhat 
below observational values.

\subsection{Comparison of Model Results with Observed Abundances in TMC-1 (CP)}
\label{compobs}
In Table \ref{table2}, the abundances of species that agree to within one 
order of magnitude at the time of best overall agreement are in normal type, 
those predicted to be more than one order of magnitude too large are in bold 
italic type and those predicted to be more than one 
order of magnitude too small are in Roman bold type.
It is immediately apparent that the inclusion of anions in both networks 
does not have a dramatic effect on the agreement with observation.  
Specifically, removing the anions from Rate06 reduces the percentage of molecular
abundances that lie within one order of magnitude of those observed from 
$\sim 66\%$ (35 out of 53 species) to $\sim 63\%$ (32 out of 51 species).
The improvement when anions are included is minor but noticeable. 
   
Important molecules for which the inclusion of anions improves agreement with 
observation are HC$_5$N, HC$_7$N and HC$_9$N (as discussed in detail in 
\S\ref{hcnn}), while the only molecule for which the inclusion of anions 
has a significantly detrimental effect on agreement with observation is C$_8$H.  
Both networks have difficulty reproducing the observed abundances of the
sulfur-containing molecules, H$_2$S, SO, C$_2$S, SO$_2$, and HCS$^+$, 
due in part to the choice of the low initial elemental abundance for S$^+$. 
A slightly higher choice can improve agreement strongly \citep{smi04}.  
With our current elemental abundance, SO, C$_2$S, and SO$_2$, all reach 
good agreement with observation at later times between $3 \times 10^5$ 
and $6 \times 10^5$ yr.  
Later times than the time of best agreement may pertain to cloudlets in 
TMC-1 other than the CP region, where most complex species peak \citep{park06}.
The calculated abundances of several organic molecules are much lower than that
observed, especially for CH$_3$OH, CH$_3$CCH, and CH$_2$CHCN.  
All these species are difficult to synthesize using gas-phase processes only.    
It has been thought for some time that saturated species, in particular, CH$_3$OH,
form on the surfaces of dust grains through the repeated hydrogenation of CO
\citep{tie82}, a claim backed up by some experimental evidence \citep{wat04}.  

\begin{figure}
\figurenum{5}
\includegraphics[angle=270,scale=0.6]{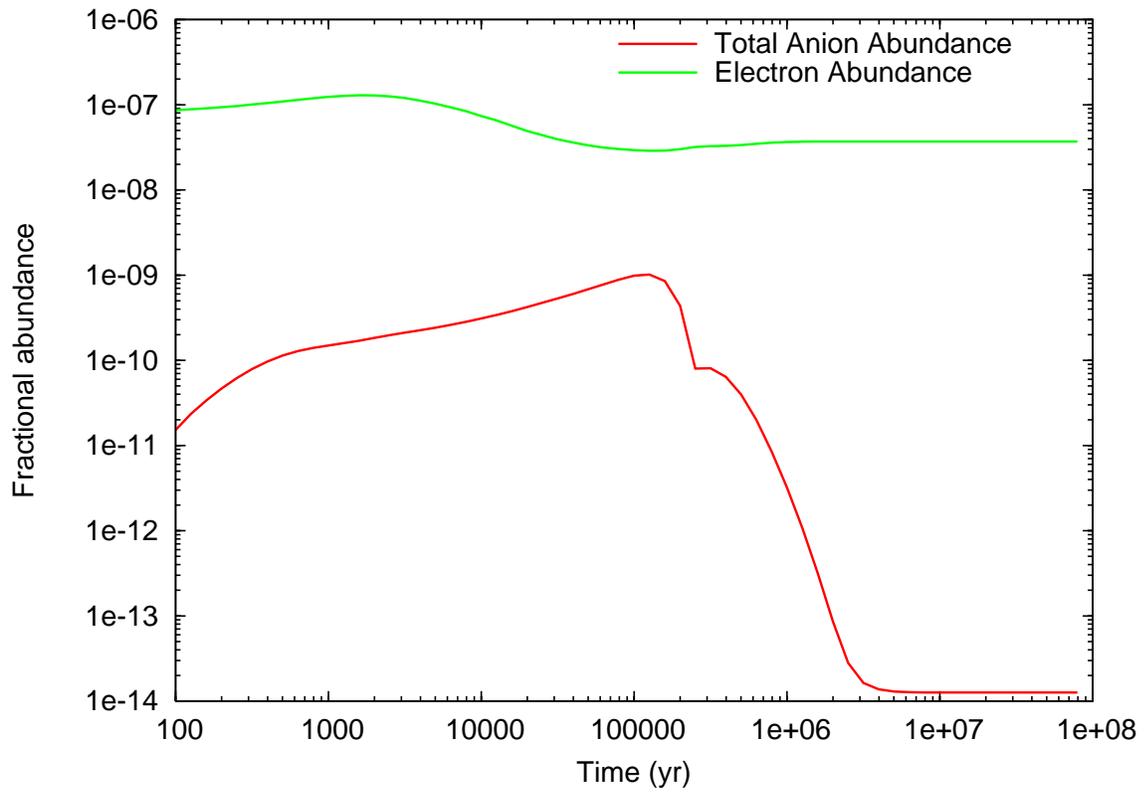}
\caption{The electron and total anionic abundances with respect to H$_2$ as functions of time computed with Rate06.}
\label{figure5}
\end{figure}

The relatively small differences observed when anions are included in our 
models are partially due to their low abundance compared with electrons, 
as shown in Figure~\ref{figure5} for Rate06.  
This is in contrast to the findings of \citet{wak08} who compute that 
after $\sim
10^2$ yr, negatively
charged small PAHs are the dominant negative charge carrier with an abundance 
almost two orders of magnitude larger than the electron abundance at steady state.  
This effect diminishes with increasing PAH size (and thus decreasing fractional
abundance).  
With PAHs  the dominant negative charge carriers,  the overall ionization 
decreases because atomic positive ions are more readily neutralized by 
PAH$^-$ than by bare electrons.  
This decrease results in increases in the abundances of positively charged 
molecular ions, which in turn lead to increases in the abundances of 
unsaturated neutral species such as the cyanopolyynes and saturated species 
such as methanol.  
Despite the increase in the abundance of methanol, however, 
the computed abundance lies well below the observed one.  
Future models should investigate the effects of adding both 
negatively charged PAHs as well as molecular anions.

\section{VARIATIONS IN INITIAL ABUNDANCES}
\label{crich}

In this work we have used oxygen-rich initial abundances with a carbon-to-oxygen 
ratio of $\sim 0.4$.  
Previous modellers \citep{ter98,rob02,smi04,wak06} have found that 
to get better agreement with observations of chemical abundances in  
TMC-1 (CP), 
a higher ratio is required, usually $\apprge 1$.  
The most successful of the models listed is the work of \citet{wak06} 
who report that with a carbon-to-oxygen ratio of 1.2, 86\% of the abundances 
can be fit if the observations are given an uncertainty of a factor of five.  
These authors use as a criterion for success an overlap between Gaussian 
distributions of modeling and observational results for each species, 
the former obtained by a Monte Carlo analysis from uncertainties in both
reaction rate coefficients and physical conditions.    

To test the effects that the addition of anions has on carbon-rich models 
of the chemistry of dark clouds, 
we have run models using both networks with a carbon-to-oxygen ratio of 1.2, 
with and without anions.
Using Rate06 without anions, at the time of best agreement, 
40 out of 51 species agree with observation to within one order of magnitude, 
an improvement similar to that found by \citet{wak06}.   
In contrast to the oxygen-rich initial conditions, we find that the addition of 
anions in the carbon-rich case worsens our agreement with 
observation for both networks.  
For Rate06 with anions,  the agreement is reduced to   
37 out of 53 species at the optimum time.
The results with the OSU network are similar.  
In the carbon-rich case, the increased synthetic power obtained upon
adding anions is magnified due to the increased carbon budget so 
that the calculated abundances of long-chain carbon molecules substantially 
overshoot the observed abundances, thus worsening the overall agreement with 
observation.     
Even in the case without anions, the better agreement obtained with 
carbon-rich abundances may be at least partially due to
an edge effect, in which much of the carbon budget ends up in the longer-chained 
species. 
Since many of these species have not been observed in TMC-1,  it is 
difficult to say whether the enhanced agreement
is real or a result of the chemistry being limited to 
molecules with 11 carbon
atoms or fewer.

\section{SUMMARY}\label{disc}
We have modelled the gas-phase chemistry of a dark cloud  
to investigate the role of molecular anions in the overall chemistry.  
We have used the simple pseudo-time-dependent gas-phase model, which can be 
criticized both for its lack of dynamics during the chemical evolution and 
for the lack of processes involving dust grains.  Nevertheless, our use of 
the simplified gas-phase model has enabled us to highlight the
direct chemical effects of the addition of molecular anions on our existing 
standard networks.

The inclusion of anions has only a slight positive effect on the success 
of both the Rate06 and OSU networks in reproducing the observed abundances 
in TMC-1 (CP) with the pseudo-time-dependent model.   
The most significant effect of the addition of anionic species
is the enhancement in the abundances of several families of carbon-chain 
molecules, namely, C$_{\mathrm{n}}$, C$_{\mathrm{n}}$H, C$_{\mathrm{n}}$H$_2$, 
C$_{\mathrm{n}}$N, and HC$_{\mathrm{n}}$N,
by providing new formation routes to the larger members of each family
via smaller molecules.  
We get much better agreement with abundances observed in TMC-1 (CP) for the
cyanopolyyne molecules, in particular, and would urge further 
experimental measurements of the rate coefficient and product channels for 
reaction~(\ref{Nreact}),  which we have shown
to be an important formation route to the HC$_{\rm n}$N molecules. 
This `catalytic' effect could be at least partially artificial 
because our chemistry is capped at molecules 
with 11 or fewer carbon atoms, so that we may be witnessing an `edge' effect whereby 
molecules move up the chain through repeated addition of a carbon atom with 
the larger molecules becoming a `sink' for carbon.  
To check if this is the case, it may be necessary to add larger 
carbon-chain species to the chemistry, similar to that done by 
\citet{mil00} for their chemical model of the shell of IRC+10216.
Credence is given to our results, however, as the peak abundances of the 
molecules remain inversely proportional to molecular size even given this 
catalytic effect.

We have also investigated the effect of using carbon-rich elemental abundances.  
Running models using both networks, with and without anions, with a
carbon-to-oxygen ratio of 1.2, we find that we get improved agreement for 
TMC-1 (CP) over the 
oxygen-rich case for the carbon-rich models without anions.  
The inclusion of anions, however,  worsens our agreement with observation 
due to the excessively enhanced
synthesis of long-chain carbon species.

Many of the large number of reactions involving anions 
that have been added to the reaction sets 
remain poorly understood.  
Although our results here show that the inclusion of 
anions does not have a dramatic effect on the chemistry, 
these conclusions must remain tentative pending 
further experimental work to determine more accurate rate
coefficients.
It is anticipated that in the near future, further anionic species will be
discovered as laboratory measurements of the rotational spectra of 
candidate molecules increase in number, and we would expect that 
existing chemical networks will be
expanded to incorporate molecular anions as standard.  
The current (anionic and non-anionic) releases of both Rate06 and the 
OSU network are available online at http://www.udfa.net/ and 
http://www.physics.ohio-state.edu/$\sim$eric/,
respectively.

\acknowledgments
{We would like to thank Valentine Wakelam and Martin Cordiner for their input and advice.  
Astrophysics at Queen's University Belfast is supported by a grant from the Science and 
Technology Facilities Council.  C. W. is supported by
scholarship from the Northern Ireland Department of Employment and Learning.  E. H.
acknowledges the support of the National Science Foundation for his research program in
astrochemistry.}

\clearpage

\clearpage

\begin{deluxetable}{lc}
\tablewidth{0pt}
\tablecaption{Initial Species and Elemental Abundances With Respect to Hydrogen Nuclear Density}
\tablehead{\colhead{Element} & \colhead{Initial Fractional Abundance}}
\startdata
H$_2$  & $5.00\times10^{-1}$ \\
He     & $1.40\times10^{-1}$ \\
C      & $7.30\times10^{-5}$ \\
N      & $2.14\times10^{-5}$ \\
O      & $1.76\times10^{-4}$ \\
S$^+$  & $2.00\times10^{-8}$ \\
Si$^+$ & $3.00\times10^{-9}$ \\
Na$^+$ & $3.00\times10^{-9}$ \\
Mg$^+$ & $3.00\times10^{-9}$ \\
Fe$^+$ & $3.00\times10^{-9}$ \\
P$^+$  & $3.00\times10^{-9}$ \\
F      & $2.00\times10^{-8}$ \\
Cl     & $3.00\times10^{-9}$ \\
\enddata
\label{table1}
\tablerefs{\citet{gra82}}
\end{deluxetable}

\begin{deluxetable}{lrcccc}
\tablecaption{Calculated Abundances of Families of Species With Respect to H$_2$  and Comparison with Observed Species  in TMC-1 (CP)
}
\tablewidth{0pt}
\tablehead{ & & \multicolumn{2}{c}{Rate06} & \multicolumn{2}{c}{OSU} \\
\colhead{Species} & \colhead{Observed\tablenotemark{a}} & \colhead{Without Anions} & 
\colhead{With Anions} & \colhead{Without Anions} & \colhead{With Anions}}
\startdata
\cutinhead{Anions}
C$_4$H$^-$     & $<$ 2.3(-12)\tablenotemark{b} & ... & 1.1(-10)& ... & 1.1(-10) \\
C$_6$H$^-$     & 1.2(-11)\tablenotemark{c}     & ... & 1.0(-10)& ... & 1.1(-10) \\
C$_8$H$^-$     & 2.1(-12)\tablenotemark{c}     & ... & 2.5(-11)& ... & 2.5(-11) \\
C$_3$N$^-$     & $<$ 7.0(-11)\tablenotemark{b} & ... & 1.6(-12)& ... & 3.6(-12) \\
\cutinhead{C$_{\mathrm{n}}$}
C$_4$          & ...\phantom{.....}  & 6.1(-09) & 3.8(-09) & 1.8(-08) & 1.6(-08) \\
C$_6$          & ...\phantom{.....}  & 6.7(-11) & 3.2(-10) & 1.1(-10) & 4.7(-10) \\
C$_8$          & ...\phantom{.....}  & 1.4(-11) & 1.4(-10) & 2.7(-11) & 2.0(-10) \\
C$_{10}$       & ...\phantom{.....}  & 8.6(-13) & 7.9(-11) & 4.3(-12) & 9.2(-11) \\
\cutinhead{C$_{\mathrm{n}}$H}
C$_4$H         & 6.1(-08)\tablenotemark{c} & 1.2(-08) & 1.2(-08)                   & 6.5(-09) & 9.8(-09)                   \\
C$_6$H         & 7.5(-10)\tablenotemark{c} & 2.1(-10) & 1.9(-09)                   & 1.8(-10) & 2.1(-09)                   \\
C$_8$H         & 4.6(-11)\tablenotemark{c} & 1.3(-11) & \textbf{\textit{6.3(-10)}} & 1.3(-11) & \textbf{\textit{6.3(-10)}} \\
C$_{10}$H      & ...\phantom{.....}        & 1.7(-13) & 1.9(-10)                   & 1.5(-12) & 1.8(-10)                   \\
\cutinhead{C$_{\mathrm{n}}$H$_2$}
C$_4$H$_2$     & ...\phantom{.....}  & 3.2(-10) & 1.2(-09) & 1.4(-10) & 1.1(-09) \\
C$_6$H$_2$     & ...\phantom{.....}  & 3.6(-11) & 4.6(-10) & 3.6(-11) & 5.9(-10) \\
C$_8$H$_2$     & ...\phantom{.....}  & 6.1(-13) & 1.6(-10) & 6.9(-13) & 2.0(-10) \\
C$_{10}$H$_2$  & ...\phantom{.....}  & 4.3(-17) & 3.2(-11) & 3.6(-17) & 3.5(-11) \\
\tablebreak
\cutinhead{C$_{\mathrm{n}}$N}
C$_3$N         & 6.0(-10)\tablenotemark{b} & 8.5(-10) & 1.1(-09) & 6.0(-10) & 1.0(-09) \\
C$_5$N         & ...\phantom{.....}        & 3.8(-11) & 9.7(-11) & 3.3(-11) & 1.5(-10) \\
C$_7$N         & ...\phantom{.....}        & 4.4(-12) & 5.0(-11) & 4.4(-12) & 6.2(-11) \\
C$_9$N         & ...\phantom{.....}        & 1.8(-13) & 1.7(-11) & 2.9(-13) & 2.0(-11) \\
\cutinhead{HC$_{\mathrm{n}}$N}
HNC$_3$        & 6.0(-11) & 2.8(-11)      & 2.4(-11) & 1.2(-10)      & 1.5(-10)      \\
HC$_5$N        & 4.0(-09) & \bf{3.6(-10)} & 4.6(-09) & \bf{2.2(-10)} & 3.1(-09)      \\
HC$_7$N        & 1.1(-09) & \bf{5.4(-11)} & 2.1(-10) & \bf{1.6(-11)} & 1.2(-10)      \\
HC$_9$N        & 4.5(-10) & \bf{3.9(-12)} & 6.8(-11) & \bf{1.9(-12)} & \bf{3.6(-11)} \\
\cutinhead{Remaining Species Observed in TMC-1}
CH           & 2.0(-08) & 5.5(-09)                   & 6.2(-09)                   & 6.6(-09)                   & 7.4(-09)                   \\
NH$_3$       & 2.0(-08) & 1.5(-08)                   & 1.3(-08)                   & 1.4(-08)                   & 1.3(-08)                   \\
OH           & 2.0(-07) & 4.0(-08)                   & 3.8(-08)                   & 2.7(-08)                   & 2.7(-08)                   \\
H$_2$O       & $\leqslant$7.0(-08) & \textbf{\textit{1.2(-06)}} & \textbf{\textit{1.4(-06)}} & \textbf{\textit{1.2(-06)}} & \textbf{\textit{1.4(-06)}} \\                            
C$_2$        & 5.0(-08) & \bf{2.3(-09)}              & \bf{2.9(-09)}              & \bf{3.7(-09)}              & \bf{3.8(-09)}              \\                  
C$_2$H       & 2.0(-08) & 1.4(-08)                   & 1.6(-08)                   & 1.1(-08)                   & 1.6(-08)                   \\
CN           & 5.0(-09) & 2.6(-08)                   & 2.8(-08)                   & 8.3(-09)                   & 9.9(-09)                   \\
HNC          & 2.0(-08) & 2.4(-08)                   & 3.3(-08)                   & 2.0(-08)                   & 2.6(-08)                   \\
HCN          & 2.0(-08) & 3.4(-08)                   & 4.8(-08)                   & 1.9(-08)                   & 2.5(-08)                   \\
HCNH$^+$     & 2.0(-09) & \bf{1.9(-10)}              & 4.5(-10)                   & 2.4(-10)                   & 4.7(-10)                   \\
CO           & 8.0(-05) & 1.0(-04)                   & 9.4(-05)                   & 9.6(-05)                   & 9.8(-05)                   \\
N$_2$H$^+$   & 4.0(-10) & 1.1(-10)                   & 1.1(-10)                   & 1.9(-10)                   & 1.7(-10)                   \\
HCO$^+$      & 8.0(-09) & 4.0(-09)                   & 4.7(-09)                   & 4.0(-09)                   & 5.0(-09)                   \\
NO           & 3.0(-08) & 3.4(-08)                   & 2.4(-08)                   & 9.1(-09)                   & 8.9(-09)                   \\
H$_2$CO      & 5.0(-08) & \bf{1.1(-09)}              & \bf{1.3(-09)}              & 8.6(-08)                   & 8.2(-08)                   \\
CH$_3$OH     & 3.0(-09) & \bf{5.1(-13)}              & \bf{4.9(-13)}              & \bf{1.2(-13)}              & \bf{1.6(-13)}              \\
O$_2$        & $\leqslant$7.7(-08) & 1.1(-07)        & 7.5(-08)                   & 7.3(-08)                   & 6.9(-08)                   \\
H$_2$S       & 5.0(-10) & \bf{7.2(-12)}              & \bf{6.9(-12)}              & \bf{9.2(-12)}              & \bf{9.1(-12)}              \\
C$_3$H       & 1.0(-08) & 1.4(-08)                   & 1.7(-08)                   & 1.0(-08)                   & 1.7(-08)                   \\
C$_3$H$_2$   & 1.1(-08) & 2.3(-08)                   & 1.5(-08)                   & 1.2(-08)                   & 1.1(-08)                   \\
CH$_2$CN     & 5.0(-09) & \bf{2.4(-10)}              & \bf{4.0(-10)}              & \bf{2.2(-10)}              & \bf{3.3(-10)}              \\
C$_2$O       & 6.0(-11) & 4.3(-11)                   & 7.5(-11)                   & 1.5(-11)                   & 1.9(-11)                   \\
CH$_3$CCH    & 6.0(-09) & \bf{4.1(-12)}              & \bf{4.2(-12)}              & \bf{4.0(-12)}              & \bf{4.6(-12)}              \\
CH$_3$CN     & 6.0(-10) & 9.3(-10)                   & 1.9(-09)                   & 4.9(-10)                   & 7.9(-10)                   \\
H$_2$CCO     & 6.0(-10) & \textbf{\textit{7.1(-09)}} & \textbf{\textit{9.4(-09)}} & \textbf{\textit{2.1(-08)}} & \textbf{\textit{2.3(-08)}} \\
CS           & 4.0(-09) & 4.9(-09)                   & 3.9(-09)                   & 7.1(-10)                   & 6.2(-10)                   \\
CH$_3$CHO    & 6.0(-10) & \bf{4.2(-11)}              & \bf{4.4(-11)}              & \bf{1.1(-12)}              & \bf{1.3(-12)}              \\
HCS$^+$      & 4.0(-10) & \bf{2.1(-12)}              & \bf{2.9(-12)}              & \bf{8.2(-13)}              & \bf{9.8(-13)}              \\
H$_2$CS      & 7.0(-10) & 1.8(-09)                   & 1.6(-09)                   & 2.4(-10)                   & 2.1(-10)                   \\
HCOOH        & 2.0(-10) & 2.0(-09)                   & \textbf{\textit{2.4(-09)}} & 7.5(-10)                   & 1.0(-09)                   \\
SO           & 2.0(-09) & \bf{5.4(-11)}              & \bf{3.8(-11)}              & \bf{6.6(-11)}              & \bf{5.7(-11)}              \\
HC$_2$NC     & 5.0(-10) & ...                        & ...                        & 2.2(-10)                   & 3.8(-10)                   \\
HNC$_3$      & 6.0(-11) & 2.8(-11)                   & 2.4(-11)                   & 1.2(-10)                   & 1.5(-10)                   \\
HC$_3$NH$^+$ & 1.0(-10) & 1.1(-11)                   & 2.3(-11)                   & 5.1(-11)                   & 1.3(-10)                   \\
C$_3$O       & 1.0(-10) & 6.0(-10)                   & 7.5(-10)                   & \textbf{\textit{1.5(-09)}} & \textbf{\textit{2.1(-09)}} \\
CH$_2$CHCN   & 4.0(-09) & \bf{5.8(-13)}              & \bf{1.3(-12)}              & \bf{1.2(-13)}              & \bf{2.0(-13)}              \\
C$_2$S       & 8.0(-09) & \bf{7.9(-10)}              & \bf{7.6(-10)}              & \bf{4.2(-10)}              & \bf{5.0(-10)}              \\
OCS          & 2.0(-09) & 4.8(-10)                   & 4.6(-10)                   & \bf{1.1(-10)}              & \bf{1.2(-10)}              \\
SO$_2$       & 1.0(-09) & \bf{5.4(-13)}              & \bf{2.8(-13)}              & \bf{8.2(-13)}              & \bf{6.7(-13)}              \\
CH$_3$C$_4$H & 4.0(-10) & \bf{2.3(-11)}              & \bf{3.4(-11)}              & \bf{3.8(-11)}              & 6.3(-11)                   \\
CH$_3$C$_3$N & 8.0(-11) & 3.0(-10)                   & 4.7(-10)                   & 9.8(-12)                   & 2.2(-11)                   \\
C$_3$S       & 1.0(-09) & 4.2(-10)                   & 4.0(-10)                   & 1.3(-10)                   & 1.6(-10)                   \\
\tableline
Agreement & ...\phantom{.....}  & 32/51 & 35/53 & 33/52 & 36/54\\ 
\enddata
\label{table2}
\tablenotetext{a}{As collated by \citet{smi04} unless otherwise stated}
\tablenotetext{b}{\citet{tha08}}
\tablenotetext{c}{\citet{bru07}}
\tablecomments{a(b) represents a $\times$ 10$^{\mathrm{b}}$}
\tablecomments{Calculated abundances refer to the time of best agreement for each model, which is  1-2 $\times 10^{5}$ yr}
\tablecomments{Normal type refers to modelled abundances which agree with
observation to within one order of magnitude, bold type refers to those 
more than an order of magnitude smaller, and bold italic type refers to
those which are more than one order of magnitude larger.}
\end{deluxetable}

\end{document}